\documentclass[reqno]{amsart}

\title[Relational lattices via duality]{Relational lattices via duality}
\author[Luigi Santocanale]{Luigi Santocanale}
\address{Luigi Santocanale\\
LIF, CNRS UMR 7279, Aix-Marseille Universit\'e}
\email{luigi.santocanale@lif.univ-mrs.fr}

\usepackage{amsmath,amssymb}
\usepackage[all]{xy}
\usepackage{wrapfig}
\RequirePackage{hyperref}

\input macros.tex
\renewenvironment{proof}{\par\noindent\emph{Proof.}~}{\par\noindent}

\begin{document}
\maketitle
\begin{abstract}
  The \emph{\nj} and the \emph{\iu} combine in different ways tables of
a relational database. Tropashko \cite{Tropashko2005} observed that
these two operations are the meet and join in a class of
lattices---called the \emph{relational lattices}---and proposed
lattice theory as an alternative algebraic approach to
databases. Aiming at query optimization, Litak et
al. \cite{LitakMHjlamp} initiated the study of the equational theory
of these lattices. We carry on with this project, making use of the
duality theory developed in \cite{San09:duality}. The contributions of
this paper are as follows. Let $A$ be a set of column's names and $D$
be a set of cell values; we characterize the dual space of the
relational lattice $\R(D,A)$ by means of a generalized ultrametric
space, whose elements are the functions from $A$ to $D$, with the
$P(A)$-valued distance being the Hamming one but lifted to subsets of
$A$. We use the dual space to present an equational axiomatization of
these lattices that reflects the combinatorial properties of these
generalized ultrametric spaces: symmetry and pairwise completeness.
Finally, we argue that these equations correspond to combinatorial properties
of the dual spaces of lattices, in a technical sense analogous of correspondence
theory in modal logic. In particular, this leads to an exact characterization of the
finite lattices satisfying these equations.

\end{abstract}

\section{Introduction}

Tropashko \cite{Tropashko2005} has recently observed that the
\emph{\nj} and the \emph{\iu}, two fundamental operations of the
relational algebra initiated by Codd \cite{Codd70}---the algebra by
which we construct queries---can be considered as the 
meet and join operations in a class of lattices, known by now as the
class of \emph{relational lattices}.
Elements of the relational lattice $\R(D,A)$ are the relations whose
variables are listed by a subset of a total set $A$ of attributes, and
whose tuples' entries are taken from a set $D$.  Roughly speaking, we
can consider a relation as a table of a database, its variables as the
columns' names, its tuples being the rows.

Let us illustrate these operations with examples. The \nj
takes two tables and constructs a new one whose columns are indexed by
the union of the headers, and whose rows are the glueings of the rows
along identical values in common columns. As we emphasize in this
paper the lattice theoretic aspects of the \nj operation, we shall
depart from the standard practice of denoting it by the symbol
$\bowtie$ and use instead the meet symbol $\land$.
\begin{center}
  \table{ll}{
    Author & Area \\ \hline
    Santocanale & Logic  \\
    Santocanale & CS   
  }
  \;$\land$\;
  \table{ll}{
    Area & Reviewer \\\hline
    CS & Turing  \\
    Logic & G\"odel}
  \;=\; \table{lll}{
    Author& Area & Reviewer \\\hline
    Santocanale & Logic & G\"odel \\
    Santocanale & CS & Turing }
\end{center}
The \iu restricts two tables to the common columns and lists all the
possible rows. The following example suggests how to construct, using
this operation, a table of users given two (or more) tables of people
having different roles.
\begin{center}
  \table{lll}{
    \multicolumn{3}{|c|}{Authors}\\
    \hline Name & Surname & Conf \\ \hline
    Luigi & Santocanale & CMCS }
  \quad$\vee$\quad
  \table{lll}{
    \multicolumn{3}{|c|}{Reviewers}\\ \hline 
    Name & Surname & Area \\\hline
    Alan & Turing & CS \\
    Kurt & G\"odel & Logic }
  \quad=\quad \table{ll}{
    \multicolumn{2}{|c|}{Users}\\ \hline 
    Name & Surname \\\hline
    Luigi & Santocanale  \\
    Alan & Turing \\
    Kurt & G\"odel }
\end{center}
Considering the lattice signature as a subsignature of the relational
algebra, Litak, Mikul{\'{a}}s and Hidders \cite{LitakMHjlamp} proposed
to study the equational theory of the relational lattices.
The capability to recognize when two queries are equivalent---that is,
a solution to the word problem of such a theory---is of course an
important step towards query optimization.

Spight and Tropashko \cite{Tropashko2008} exhibited equational
principles in a signature strictly larger than the one of lattice
theory. A main contribution of Litak et al.  \cite{LitakMHjlamp}---a
work to which we are indebted in many respects---was to show that the
quasiequational theory of relational lattices \emph{with the header
  constant} is undecidable.  The authors also proposed a base of
equations for the theory in the signature extended with the header
constant, and exhibited two non-trivial pure lattice equations holding
on relational lattices.  It was argued there that the lattice
$\R(D,A)$ arises via a closure operator on the powerset
$P(A \sqcup \expo{A}{D})$ and, at the same time, as the \Groth
construction for the functor $P(\expo{(-)}{D})$, from $P(A)^{op}$ to
$\JSL$ (the category of complete lattices and join-preserving mappings),
sending $X \subseteq A$ to $\expo{X}{D}$ and then $\expo{X}{D}$
covariantly to $P(\expo{X}{D})$.

The focus of this paper is on the pure lattice signature.  We tackle
the study of the equational theory of relational lattices in a
coalgebraic fashion, that is, by using the duality theory developed in
\cite{San09:duality} for finite lattices and here partially extended
to infinite lattices. Let us recall some key ideas from the theory,
which in turn relies on 
Nation's representation Theorem \cite[\S 2]{Nation90}.
For a complete lattice $L$, a \jc of $x \in L$ is a subset
$Y \subseteq L$ such that $x \leq \bv Y$.  A lattice is
\emph{\pperfect} if it is a complete spatial lattice and every \jc of
a \cjirr element refines to a minimal one---see
Section~\ref{sec:defselconcepts} for a complete definition. Every
finite lattice is \pperfect; moreover, relational lattices are
\pperfect, even when they are infinite. This property,
i.e. \pperfect{ness}, allows to define the dual structure of a lattice
$L$, named the OD-graph in \cite{Nation90}. This is the triple
$\langle \Ji(L),\leq,\mcovered \rangle$ with $\Ji(L)$ the set of
\cjirr elements, $\leq$ the restriction of the order to $\Ji(L)$, and
the relation $j \mcovered C$ holds when $j \in \Ji(L)$,
$C \subseteq \Ji(L)$, and $C$ is a \mjc of $j$.  The original lattice
$L$ can be recovered up to isomorphism from its OD-graph as the
lattice of closed downsets of $\Ji(L)$---where a downset
$X \subseteq \Ji(L)$ is closed if $j \mcovered C \subseteq X$ implies
$j \in X$.

\medskip

We characterize the OD-graph of the lattice $\R(D,A)$ as follows.
Firstly recall from \cite{LitakMHjlamp} that we can identify \cjirr
elements of $\R(D,A)$ with elements of the disjoint sum
$A \sqcup \expo{A}{D}$.  The order on \cjirr elements is trivial,
i.e. it is the equality. All the elements of $A$ are \jp, whence the
only \mjc of some $a \in A$ is the singleton $\set{a}$.
The \mjc{s} of elements in $\expo{A}{D}$ are described via an
ultrametric distance valued in the \jsl $P(A)$; this is, morally, the
Hamming distance, $\d(f,g) = \set{x \in A \mid f(x) \neq g(x)}$.
Whenever $f,g \in \expo{A}{D}$ we have
$f \mcovered \d(f,g) \cup \set{g}$ and these are all the \mjc{s} of
$f$.

As in correspondence theory for modal logic, the combinatorial
structure of the dual spaces is an important source for discovering
axioms/equations that uniformly hold in a class of models. For
relational lattices, most of these combinatorial properties stem from
the structure of the ultrametric space $(\AD, \d)$. When we firstly
attempted to show that equations \AxRLone and \AxRLtwo from
\cite{LitakMHjlamp} hold in relational lattices using duality, we
realized that the properties necessary to enforce these equations were
the following:
\begin{enumerate}[{P}1.]
\item Every non-trivial \mjc contains at most one  \jirr element which
  is not \jp.
  \label{P:Unjp}
  \end{enumerate}
  Moreover, the generalized ultrametric space $(\AD,\d)$ is
  \begin{enumerate}[{P}1.]
    \setcounter{enumi}{1}
  \item \emph{symmetric}, i.e. $\d(f,g) = \d(g,f)$, for each $f,g \in \AD$,
    \label{P:Sym}
  \item \emph{\Pc}: if $\d(f,g) \subseteq X \cup Y$, then
    $\d(f,h) \subseteq X$ and $\d(h,g) \subseteq Y$ for some
    $h \in \AD$.
    \label{P:BC}  
\end{enumerate}
Various notions of completeness for generalized ultrametric spaces are
discussed in \cite{Ackerman2013}.
At first we called \PC the \BCMp of $(\AD,\d)$. Indeed, it is
equivalent to saying that the functor
$P(\expo{(-)}{D}) : P(A)^{op}\!\rto \JSL$ mentioned above sends a
pullback square (i.e., a square of inclusions with objects
$X \cap Y,X,Y,Z$) to a square satisfying the Beck-Chevalley
condition. As the property implies that a collection of congruences of
\jsl{s} commute, it is also a sort of \malcev condition.

We show with Theorem~\ref{thm:corrunjp} that property \Pref{Unjp} of
an OD-graph is definable by an \identity that we name \Unjp. We
investigate the deductive strength of this \identity and show in
particular that \AxRLtwo is derivable from \Unjp, but not the
converse.

In presence of \Pref{Unjp}, symmetry and \PC can also be
understood as properties of an OD-graph. Symmetry is the following
property: if $k_{0} \mcovered C \cup \set{k_{1}}$ with $k_{1}$ not
\jp, then $k_{1} \mcovered C \cup\set{k_{0}}$. \PPC can be read as
follows: if $k_{0} \mcovered C_{0} \cup C_{1} \cup \set{k_{2}}$ with
$k_{2}$ not \jp and $C_{0},C_{1},\set{k_{2}}$ pairwise disjoint, then
$k_{0} \mcovered C_{0} \cup \set{k_{1}}$ and
$k_{1} \mcovered C_{1} \cup \set{k_{2}}$ for some \cjirr element
$k_{1}$.

We exhibit in Section~\ref{sec:othereqs} three equations valid on
relational lattices and characterize, via a set of properties of their
OD-graphs, the \pperfect lattices satisfying \Unjp and these
equations. We propose these four equations as an axiomatization of the
theory of relational lattices that we call \Axioms.  The main result
of this paper, Theorem~\ref{thm:atomisticcharacterization}, sounds as
follows. If we restrict to \emph{finite} lattices that are
\emph{atomistic}---that is, lattices in which any element is the join
of the atoms below it, so the order on \jirr elements in the dual
space is trivial---then a lattice satisfies \Axioms if and only if its
OD-graph is symmetric and \Pc, in the sense just explained.

We can build lattices similar to the relational lattices from
$P(A)$-valued ultrametric spaces. It is tempting to look for further
equations so to represent the OD-graph of finite atomistic lattices
satisfying these equations as $P(A)$-valued ultrametric
space. Unfortunately this is not possible, since a key property of the
OD-graph of lattices of ultrametric spaces---the ones ensuring that
the distance function is well defined---is not definable by lattice
equations. Yet Theorem~\ref{thm:atomisticcharacterization} also
exhibits a deep connection between the OD-graph of finite atomistic
lattices satisfying \Axioms and the frames of the commutator logic
$[\Sfive]^{A}$, see \cite{Kurucz2007}. Considering the complexity of
the theory of combination of modal logics,
Theorem~\ref{thm:atomisticcharacterization} can be used to foresee and
shape future researches. For example, we shall discuss in
Section~\ref{sec:conclusions} how to derive undecidability results
from the correspondent ones in multidimensional modal logic. In
particular, a refinement of the main result of Litak et al.
\cite[Corollary~4.8]{LitakMHjlamp} can be derived.

The paper is structured as follows. We introduce in
Section~\ref{sec:defselconcepts} the notation as well as the least
lattice theoretic tools that shall allow the reader to go through the
paper. In Section~\ref{sec:rellattices} we describe the relational
lattices, present some known results in the literature, and give a
personal twist to these results. In particular, we shall introduce
semidirect products of lattices, ultrametric spaces as a tool for
studying relational lattices, emphasize the role of the \BCp in the
theory.  In Section~\ref{sec:mjcovers} we characterize the OD-graphs
of relational lattices. In Section~\ref{sec:unjp} we present our
results on the equation \Unjp. In Section~\ref{sec:othereqs}, we
describe our results relating equations valid on relational lattices
to symmetry and \PC.  In the last Section we discuss the results
presented as well as ongoing researches, by the author and by other
researchers, trace a road-map for future work.

\section{Some elementary lattice theory}
\label{sec:defselconcepts}

A \emph{lattice} is a poset $L$ such that every finite non-empty
subset $X \subseteq L$ admits a smallest upper bound $\bv X$ and a
greatest lower bound $\bigwedge X$. We assume a minimal knowledge of
lattice theory---otherwise, we invite the reader to consult a standard
monograph on the subject, such as \cite{DP02} or \cite{GLT2}. The
technical tools that we use may be found in the monograph \cite{FJN},
that we also invite to explore. A lattice can also be understood as a
structure $\A$ for the functional signature $(\vee,\land)$, such that
the interpretations of these two binary function symbols both give
$\A$ the structure of an idempotent commutative semigroup, the two
semigroup structures being tied up by the absorption laws
$x \land (y \vee x) = x$ and $x \vee (y \land x) = x$.
Once a lattice is presented as such structure, the order is recovered
by stating that $x \leq y$ holds if and only if $x \land y= x$.

A lattice $L$ is \emph{complete} if any subset $X \subseteq L$ admits
a smallest upper bound $\bv X$. It can be shown that this condition
implies that any subset $X \subseteq L$ admits a greatest lower bound
$\bigwedge X$. A complete lattice is \emph{bounded}, since
$\bot := \bigvee \emptyset$ and $\top := \bigwedge \emptyset$ are
respectively the least and greatest elements of the lattice.

A \emph{closure operator} on a complete lattice $L$ is an
order-preserving function $j : L \rto L$ such that $x \leq j(x)$ and
$j^{2}(x) = j(x)$, for each $x \in L$.  We shall use $\Clop(L)$ to
denote the poset of closure operators on $L$, under the pointwise
ordering. It can be shown that $\Clop(L)$ is itself a complete
lattice. If $j \in \Clop(L)$, then the set $L/j$ of fixed points of
$j$ is itself a complete lattice, with
$\bigwedge_{L/j} X = \bigwedge_{L} X$ and
$\bigvee_{L/j} X = j(\bigvee_{L} X)$. For the correspondence between
closure operators and congruences in the category of complete
join-semilattices, see \cite{joyaltierney}.

Let $L$ be a complete lattice. An element $j \in L$ is said to be
\emph{\cjirr} if $j = \bv X$ implies $j \in X$, for each
$X \subseteq L$; the set of \cjirr element of $L$ is denoted here
$\Ji(L)$. A complete lattice is \emph{spatial} if every element is the
join of the \cjirr elements below it.  An element $j \in \Ji(L)$ is
said to be \emph{\jp} if $j \leq \bv X$ implies $j \leq x$ for some
$x \in X$, for each finite subset $X$ of $L$. We say that
$j \in \Ji(L)$ is \emph{\njp} if it is not \jp. An \emph{atom} of a
lattice $L$ is an element of $L$ such that $\bot$ is the only element
strictly below it. A spatial lattice is \emph{atomistic} if every
element of $\Ji(L)$ is an atom.

For $j \in \Ji(L)$, a \emph{join-cover} of $j$ is a subset $X
\subseteq L$ such that $j \leq \bv X$. For $X, Y \subseteq L$, we say
that $X$ \emph{refines} $Y$, and write $X \refines Y$, if for all $x
\in X$ there exists $y \in Y$ such that $x \leq y$. A join-cover $X$
of $j$ is said to be \emph{minimal} if $j \leq \bv Y$ and $Y \refines
X$ implies $X \subseteq Y$; we write $j \mcovered X$ if $X$ is a \mjc
of $j$. In a spatial lattice, if $j \mcovered X$, then $X \subseteq
\Ji(L)$. If $j \mcovered X$, then we say that $X$ is a
\emph{non-trivial} \mjc of $j$ if $X\neq \set{j}$. It is common to use
the word perfect for a lattice which is both spatial and dually
spatial. We need here something different:
\begin{definition}
  A complete lattice is \emph{\pperfect} if it is spatial
  and for each $j \in \Ji(L)$ and $X \subseteq L$, if $j \leq \bv X$, then $Y
  \refines X$ for some $Y$ such that $j \mcovered Y$.
  The \emph{OD-graph} of a \pperfect lattice $L$ is the structure $\langle
  \Ji(L),\leq,\mcovered \rangle$.
\end{definition}
That is, in a \pperfect lattice every cover refines to a minimal
one. Notice that every finite lattice is \pperfect.  If $L$ is a
\pperfect lattice, then we say that $X \subseteq \Ji(L)$ is
\emph{closed} if it is a downset and $j \mcovered C \subseteq X$ implies $j \in
X$. As from standard theory, the mapping $X \mapsto \bigcap \set{ \,Y
  \subseteq \Ji(L) \mid X \subseteq Y, Y \text{ is closed} }$ defines
a closure operator whose fixed points are exactly the closed subsets
of $\Ji(L)$.  The interest of considering \pperfect lattices stems from
the following representation Theorem. 
\begin{theorem}[Nation \cite{Nation90}]
  \label{thm:nation}
  Let $L$ be a
  \pperfect lattice and 
  let $\L(\Ji(L),\leq,\mcovered)$ be the lattice of
  closed subsets of $\Ji(L)$. The mapping $l \mapsto \set{j \in
    \Ji(L) \mid j \leq l}$ is a lattice isomorphism from $L$ to
  $\L(\Ji(L),\leq,\mcovered)$.
\end{theorem}
It was shown in \cite{San09:duality} how to extend this representation
theorem to a duality between the category of finite lattices and the
category of OD-graphs.  The following Lemma shall be repeatedly used
in the proofs of our statements.
\begin{lemma}
  \label{lemma:elofmjc}
  Let $L$ be a \pperfect lattice, let $j \mcovered C$ and $k \in C$.
  If $j \leq \bv D$ with $D \refines \set{\bv (C \setminus \set{k}),
    k}$, then $k \in D$. In particular, if $k' < k$, then $\set{\bv C
    \setminus \set{k},k'}$ is not a cover of $j$.
\end{lemma}

\section{The relational lattices $\R(D,A)$}
\label{sec:rellattices}

In this Section we define relational lattices, recall some known
facts, and develop then some tools to be used later, semidirect
products of lattices, generalized ultrametric spaces, a precise
connection to the theory of combination of modal logics (as well as
multidimensional modal logic and relational algebras).

Let $A$ be a collection of attributes (or column names) and let $D$ be
a set of cell values. A \emph{relation} (or, more informally, a
\emph{table}) on $A$ and $D$ is a pair $(X,T)$ where $X \subseteq A$
and $T \subseteq \expo{X}{D}$; $X$ is the header of the table while
$T$ is the collection of rows. Elements of the relational lattice
$\R(D,A)$ are relations on $A$ and $D$.
  
Before we define the \nj, the \iu operations, and the order on
$\R(D,A)$, let us recall a few key operations. If
$X \subseteq Y \subseteq A$ and $f \in \expo{Y}{D}$, then we shall use
$f \restr[X] \in \expo{X}{D}$ for the restriction of $f$ to $X$; if
$T \subseteq \expo{Y}{D}$, then $T \rrestr[X]$ shall denote projection
to $X$, that is, the direct image of $T$ along restriction,
$T \rrestr[X] := \set{ f \restr[X] \mid f \in T}$; if
$T \subseteq \expo{X}{D}$, then $i_{Y}(T)$ shall denote
cylindrification to $Y$, that is, the inverse image of restriction,
$i_{Y}(T) := \set{ f \in \expo{Y}{D} \mid f_{\restriction X} \in T}$.
Recall that $i_{Y}$ is right adjoint to $\rrestr[X]$.  With this in
mind, the \nj and the inner union of tables are respectively described
by the following formulas:
\begin{align*}
  (X_{1},T_{1}) \land (X_{2},T_{2})
  & := (X_{1} \cup X_{2},T)  \\
  \text{where }T & = \set{f \mid f \restr[X_{i}] \in T_{i}, i = 1,2} =
  i_{X_{1} \cup X_{2}}(T_{1}) \cap i_{X_{1} \cup
    X_{2}}(T_{2})\,,  \\
  (X_{1},T_{1}) \vee (X_{2},T_{2})
  & := (X_{1} \cap X_{2},T) \\
  \text{where }T & = \set{f \mid \exists i\in \set{1,2},\exists
    g \in T_{i} \tst g\, \restr[X_{1} \cap X_{2}] = f} \\
  & = T_{1} \rrestr[X_{1} \cap X_{2}] \cup \,T_{2} \rrestr[X_{1} \cap
  X_{2}]\,.
\end{align*}
The order is then given by 
\begin{align*}
  (X_{1},T_{1}) & \leq (X_{2},T_{2}) \qquad \tiff \qquad X_{2} \subseteq X_{1}
  \tand T_{1} \rrestr[X_{2}] \subseteq T_{2}\,.
\end{align*}

It was observed in \cite{LitakMHjlamp} that $\R(D,A)$ arises---as a
category with at most one arrow between two objects---via the
Grothendieck construction for the functor sending $X \subseteq A$
contravariantly to $\expo{X}{D}$ and then $\expo{X}{D}$ covariantly to
$P(\expo{X}{D})$.
Let us record the following important property:
\begin{lemma}
  \label{lemma:BC}
  The image of a pullback square by the functor $P(\expo{(-)}{D}) :
  P(A)^{op} \rto \JSL$ satisfies the Beck-Chevalley property.
\end{lemma}

$$
\xymatrix{
  P(\expo{X_{1} \cap X_{2} }{D}) \ar@/^1em/[rr]^{i_{X_{2}}}
  & & P(\expo{X_{2}}{D})
  \ar[ll]^{\rrestr[X_{1} \cap X_{2}]}\\
  P(\expo{X_{1}}{D}) 
  \ar@/^1em/[rr]^{i_{X_{3}}}
  \ar[u]^{\rrestr[X_{1} \cap X_{2}]} & & P(\expo{X_{3}}{D})\ar[u]^{\rrestr[X_{2}]}
  \ar[ll]^{\rrestr[X_{1}]} }
$$
The above statement means that if we apply the functor to inclusions
of the form $X_{1} \cap X_{2} \subseteq X_{i} \subseteq X_{3}$, $i =
1,2$, then the two possible diagonals in the diagram above,
$\rrestr[X_{2}]\circ \,i_{X_{3}}$ and $i_{X_{2}} \circ \rrestr[X_{1}
\cap X_{2}]$, are equal.  The \BCp is a consequence of the glueing
property of functions: \emph{if $f \in \expo{X_{2}}{D}, g \in
  \expo{X_{1}}{D}$ and $f \restr[X_{1} \cap X_{2}] = g\, \restr[X_{1}
  \cap X_{2}]$, then there exists $h \in \expo{X_{3}}{D}$ such that
  $h\,\restr[X_{2}] = f$ and $h\, \restr[X_{1}] = g$.}

\medskip

We can recast the previous category-theoretic observations in an
algebraic framework.  An \emph{action} of a complete lattice $L$ over
a complete lattice $M$ is a 
monotonic mapping $\pos : L \rto \Clop(M)$, thus sending $X \in L$ to
a closure operator $\pos[X]$ on $M$. Given such an action, if we
define $j(X,T) := (X,\pos[X]T)$,
then $j(X,T)$ is a closure operator on the product $L \times M$. In
particular, 
the set of $j$-fixed points,
$L \sdp M := \set{(X,T) \in L \times M \mid \pos[X]T = T }$, is itself
a complete lattice, where the meet coincides with the one from
$L \times M$, while the join is given by the formula
$(X_{1},T_{1}) \vee_{L \sdp M} (X_{2},T_{2}) := (X_{1} \vee
X_{2},\pos[X_{1} \vee X_{2}](T_{1} \vee T_{2}))$.
We call $L \sdp M$ the \emph{semidirect product} of $L$ and $M$ via
$j$. 
The naming is chosen here after the semidirect product of groups,
which is a similar instance of the \Groth construction.
Given such an action, the correspondence $X \mapsto M/\pos[X]$
gives rise to a covariant functor from $L$ to the category $\JSL$, so
that
it makes sense to ask when the \BCp holds, as in Lemma~\ref{lemma:BC}.
This happens---and then we say that an action $\pos$ satisfies the \BC
property---exactly when
\begin{align}
  \label{eq:BCLPAR}
  \pos[X_{1} \vee X_{2}] T & = \pos[X_{1}]\pos[X_{2}]T\,, & \text{for
    each $X_{1},X_{2} \in L$ and $T \in M$.}
\end{align}
Notice that the identity
$\pos[X_{1}]\pos[X_{2}]T = \pos[X_{2}]\pos[X_{1}]T$ is a consequence
of \eqref{eq:BCLPAR}. As these closure operators correspond to
congruences of complete \jsl{s}, we also think of the \BCp as a form
of Malcev property, stating that a collection of congruences (thought
as binary relations) pairwise commute (w.r.t composition of
relations).

\medskip

\paragraph{Relational lattices from ultrametric spaces.}
Let us come back to the lattice $\R(D,A)$. Define on the set
$\expo{A}{D}$ the following $P(A)$-valued ultrametric
distance:
\begin{align*}
  \d(f,g) & := \set{ x \in A \mid f(x) \neq g(x) }\,.
\end{align*}
Thus $\d(f,f) \subseteq \emptyset$ and
$\d(f,g) \subseteq \d(f,h) \cup \d(h,g)$ for any
$f,g,h \in \expo{A}{D}$, making $(\expo{A}{D},\d)$ into a generalized
metric space in the sense of \cite{lawvere}.\footnote{This is a
  lifting of the Hamming distance to subsets. Yet, in view of
  \cite{PriessCrampeRibenboim1995} and of their work on generalized
  ultrametric spaces, such a distance might reasonably tributed to
  Priess-Crampe and Ribenboim.}  With respect to the latter
work---where axioms for the distance are those of a category enriched
over $(P(A)^{op},\emptyset,\cup)$---for $f, g \in \expo{A}{D}$ we also
have that $\d(f,g) = \emptyset$ implies $f = g$ and symmetry,
$\d(f,g) = \d(g,f)$.
We can define then an action of $P(A)$ on $P(\expo{A}{D})$: 
\begin{align}
  \label{def:distance}
  \pos[X]T & = \set{ f \in \expo{A}{D} \mid \exists g \in T \tst
    \delta(f,g) \subseteq X}\,.
\end{align}
We can now restate (and refine) Lemma 2.1 from
\cite{LitakMHjlamp}---which constructs the lattice $\R(D,A)$ via a
closure operator on $P(A + \expo{A}{D}$)---as follows:
\begin{theorem}
  \label{thm:sdp}
  The correspondence sending $(X,T)$ to $(A \setminus X,i_{A}(T))$ is an
  isomorphism bewteen the relational lattice $\R(D,A)$ and
  $P(A) \sdp P(\expo{A}{D})$.
\end{theorem}
The action defined in \eqref{def:distance} satisfies the identity
\eqref{eq:BCLPAR}. As a matter of fact, \eqref{eq:BCLPAR} is
equivalent to \emph{pairwise completeness} of $(\expo{A}{D},\delta)$
as an ultrametric space, see \cite{Ackerman2013}, namely the following
property:
\begin{align}
  \notag
  \smalllhs[5mm]{\text{\emph{if $\d(f,g) \subseteq X_{1} \cup X_{2}$,}}} \\
  &\qquad\text{\emph{then there exists $h$
      such that $\d(f,h) \subseteq X_{1}$ and $\d(h,g) \subseteq
      X_{2}$}}\,.
  \label{cond:BC}
\end{align}
It is easily verified that \eqref{cond:BC} is yet another spelling of
the glueing property of functions.

Observe that, given any generalized ultrametric space $(F,\d)$ whose
distance takes values in $P(A)$, equation \eqref{def:distance}---with
$\expo{A}{D}$ replaced by $F$---defines an action of $P(A)$ on $P(F)$.
The lattice $P(A) \sdp P(F)$ shall have similar properties to those of
the lattices $\R(D,A)$ and will be useful when studying the variety
generated by the relational lattices.
As an example, we construct \emph{typed relational lattices},
i.e. lattices of relations where each column has a fixed type. To this
goal, fix a surjective mapping $\pi : D \rto A$. For each $a \in A$,
we think of the set $D_{a} = \pi^{-1}(a)$ as the type of the attribute
$a$.  Let $S(\pi)$ be the set of sections of $\pi$, that is,
$s \in S(\pi)$ if and only if $s(a) \in D_{a}$, for each $a \in A$.
Notice that $(S(\pi),\d)$ is a pairwise complete sub-metric space of
$(\expo{A}{D},\d)$. The lattice $\R(\pi) := P(A) \sdp P(S(\pi))$ is
the typed relational lattice.  It can be shown that relational
lattices and typed relational lattices generate the same variety.

\paragraph{Relational lattices from multidimensional modal logics.}
In order to illustrate and stress the value of identity
\eqref{eq:BCLPAR}, i.e. of the \BCMp, we derive next a useful formula
for computing the join of two tables under the
$P(A) \sdp P(\expo{A}{D})$ representation.
\begin{align*}
  (X_{1},T_{1}) \vee (X_{2},T_{2}) & = (X_{1}\cup X_{2},\pos[X_{1}\cup
  X_{2}](T_{1} \cup T_{2}))\\
  & = (X_{1}\cup X_{2},\pos[X_{1}\cup X_{2}]T_{1} \cup \pos[X_{1}\cup
  X_{2}]T_{2}) \tag*{since the modal operators
    $\pos[X]$ are normal, in
    the usual sense of modal logic,} \\
  & = (X_{1}\cup X_{2},\pos[X_{2}]\pos[X_{1}]T_{1} \cup
  \pos[X_{1}]\pos[X_{2}]T_{2})
  \tag*{by  \eqref{eq:BCLPAR}} \\
  & = (X_{1}\cup X_{2},\pos[X_{2}]T_{1} \cup \pos[X_{1}]T_{2})
  \tag*{since $\pos[X_{1}]T_{1} = T_{1}$ and $\pos[X_{2}]T_{2} =
    T_{2}$.}
\end{align*}
Theorem \ref{thm:sdp} suggests that a possible way to study the
equational theory of the lattices $\R(D,A)$ is to interpret the
lattice operations in a two sorted modal logic, where the modal
operators are indexed by the first sort and act on the second.  It is
easily recongnized that each modal operator satisfies the \Sfive
axioms, while equation~\eqref{eq:BCLPAR} implies that, when $A$ is
finite, each modal operator $\pos[X]$ is determined by the modal
operators of the form $\pos[a]$ with $a$ an atom below $X$. That is,
the kind of modal logic we need to interpret the lattice theory is the
commutator logic
$[\Sfive]^{n} = [\underbrace{\Sfive,\ldots
  ,\Sfive}_{\text{$n$-times}}]$, with $n = \card A$, see
\cite[Definition 18]{Kurucz2007}.

\section{Minimal join-covers in $\R(D,A)$}
\label{sec:mjcovers}

The lattices $\R(D,A)$ are \pperfect, even when $A$ or $D$ is an
infinite set. The \cjirr elements were characterized in
\cite{LitakMHjlamp} together with the meet-irreducible elements and
the canonical context (see \cite[Chapter 3]{DP02} for the definition
of canonical context). If we stick to the representation given in
Theorem~\ref{thm:sdp}, the \cjirr elements are of the form
$\JJP{a} = (\set{a},\emptyset)$ and $\JJP{f} = (\emptyset,\set{f})$.
We can think of $\JJP{a}$ as an empty named column, while $\JJP{f}$ is
an everywhere defined row. They are all atoms, so that, in particular,
we shall not be concerned with the restriction of the order to
$\Ji(\R(D,A))$ (since this order coincides with the equality). In
order to characterize the OD-graph of the lattice $\R(D,A)$, we only
need to characterize the \mjc{s}. Taking into account that if an
element $j \in \Ji(L)$ is \jp, then it has just one \mjc, the
singleton $\set{j}$, the following Theorem achieves this goal.
\begin{theorem}
  The lattices $\R(D,A)$ are atomistic \pperfect lattices.  As a
  matter of fact, every element $\JJP{a}$, $a \in A$, is \jp; for
  $f \in \expo{A}{D}$, the \mjc{s} of $\JJP{f}$ are of the form
  \begin{align*}
    \JJP{f} & \leq \textstyle{ \bv_{a \in \d(f,g)} \JJP{a}} \vee
    \JJP{g}\,,
    \qquad\text{for  $g \in \expo{A}{D}$.}
  \end{align*}
\end{theorem}
The proof of this statement is almost straightforward, given the
characterization of $\R(D,A)$ as the semidirect product
$P(A) \sdp P(\AD)$ and the definition of the closure operators given
with equation~\eqref{def:distance}. For this reason, we skip it.

In particular, \emph{every \mjc contains at most one \njp element}. In
view of Theorem~\ref{thm:nation}, we obtain a more precise description
of the closure operator described in \cite[Lemma 2.1]{LitakMHjlamp}
that gives rise to relational lattices.
\begin{corollary}
  The relational lattice $\R(D,A)$ is isomorphic to the lattice of
  closed subsets of $A \sqcup \expo{A}{D}$, where a subset $X$ is
  closed if $\d(f,g) \cup \set{g} \subseteq X$ implies $f \in X$.
\end{corollary}
In order to ease the reading, we shall use in the rest of this paper
the same notation for a \cjirr element of $\R(D,A)$ and an element of
$A \sqcup \AD$. This is consistent with the above Corollary, as under
the isomorphism we have $\JJP{a} = \set{a}$ and $\JJP{f} =
\set{f}$. Thus $\JP{a}$ shall stand for $\JJP{a}$, and $\JP{f}$ for
$\JJP{f}$.

In a spatial lattice $L$ (thus in a relational lattice), an inequation
$s \leq t$ holds if and only if $j \leq s$ implies $j \leq t$, for
each $j \in \Ji(L)$---thus we shall often consider inequations of the
form $j \leq s$ with $j \in \Ji(L)$. When $L = \R(D,A)$, the
characterization of \mjc leads to the following principle that we
shall repeatedly use:
\begin{align}
  \label{eq:jirrmjc}
  f & \leq x_{1} \vee \ldots \vee x_{n}  \tiff
  \d(f,g) \cup \set{g} \refines \set{x_{1},\ldots ,x_{n}},\text{ for
    some $g \in \AD$}.
\end{align}

\section{Uniqueness of \njp elements}
\label{sec:unjp}

For an \emph{inclusion} we mean a pair $(s,t)$ of terms (in the
signature of lattice theory) such that the equation $t \vee s = s$
(i.e. the inequality $t \leq s$) is derivable from the usual axioms of
lattices. Thus, the equality $s = t$ reduces to the inequality $s \leq
t$. We write $s \leq t$ for a lattice inclusion and say it holds in a
lattice if the identity $s = t$ holds in that lattice. Next, let us set
\begin{align*}
    \ld[u] := u_{0} \land (u_{1} \vee u_{2})\,, \quad
    \rd[u] := (u_{0} \land u_{1}) \vee (u_{0} \land u_{2})\,,
\end{align*}
so $\ld[u] \leq \rd[u]$ is (an inclusion equivalent to) the usual
distributive law. Consider the following inclusion:
\begin{align}
  \label{eq:Unjp}
  \tag{Unjp} \smalllhs[6mm]{
    x \land (\ld[y] \vee \ld[z] \vee w) } \\
  \notag & \leq (x \land (\rd[y] \vee \ld[z] \vee w)) \vee (x \land
  (\ld[y] \vee \rd[z] \vee w))\,.
\end{align}

\begin{theorem}
  \label{thm:corrunjp}
  The inclusion \Unjp holds on relational lattices. As a matter of
  fact, \Unjp holds in a \pperfect lattice if and only if every
  minimal join-cover contains at most one non-join-prime element.
\end{theorem}
\begin{proof}
  Let us prove the first statement. To this goal, it will be enough to
  argue that any \jirr element below the \lhs of the inclusion is
  below its \rhs. Let $k$ be such a \jirr element.  It is not
  difficult to see that if $k$ is \jp, then $k$ is also below the \rhs
  of the inclusion. Suppose then that $k$ is \njp, whence $k = \JP{f}$
  for some $f \in \AD$. From $\JP{f} \leq \ld[y] \vee \ld[z] \vee w$
  and \eqref{eq:jirrmjc}, it follows that
  there exists $g \in \AD$ such that $\JP{\d(f,g)} \cup
  \set{\JP{g}}\refines \set{\ld[y],\ld[z],w}$. In particular,
  $\set{\JP{g}}\refines \set{\ld[y],w}$ or $\set{\JP{g}}\refines
  \set{\ld[z],w}$. We firstly suppose that the last case holds.
  If $a \in \delta(f,g)$ and $\JP{a} \leq \ld[y]= y_{0} \land (y_{1}
  \vee y_{2})$, then $\JP{a} \leq (y_{0} \land y_{1}) \vee (y_{0}\land
  y_{2}) = \rd[y]$, since $\JP{a}$ is \jp. It follows that
  $\JP{\delta(f,g)} \cup \set{\JP{g}}\refines \set{\rd[y],\ld[z],w}$,
  whence $\JP{f} \leq x \land (\ld[y] \vee \ld[z] \vee w)$. If
  $\set{\JP{g}} \refines \set{\ld[y],w}$, then we conclude similarly
  that $\JP{f} \leq x \land (\ld[y] \vee \rd[z] \vee w)$.  Whence $k$
  is below the \rhs of this inclusion, and the inclusion holds since
  $k$ was arbitrary.

  We leave the reader to generalize the argument above so to prove
  that if a \pperfect lattice is such that every \mjc has at most one
  \njp element, then \Unjp holds. For the converse we argue as
  follows.

  Let $L$ be a \pperfect lattice, let $k_{x} \in \Ji(L)$, $C_{x} \subseteq
  \Ji(L)$ with $k_{x} \mcovered C_{x} $, and suppose that $k_{y},k_{z}
  \in C_{x}$ are distinct and \njp.
  For $u \in \set{y,z}$, since $k_{u}$ is \njp, there is a non-trivial
  \mjc $k_{u} \mcovered C_{u}$; as every non-trivial \mjc has at least
  two elements, let $C_{u,1},C_{u,2}$ be a partition of $C_{u}$ such
  that $C_{u,i}\neq \emptyset$ for each $i = 1,2$.
  
  We construct a valuation which fails \Unjp.
  Let $x := k_{x}$, $y_{0} := k_{y}$, $z_{0} := k_{z}$, $w := \bigvee
  (C_{x} \setminus \set{k_{y},k_{z}})$ and, for $u \in \set{y,z}$ and
  $i =1,2$, let $u_{i} := \bigvee C_{u,i}$.
  The \lhs of the \Unjp evaluates to $k_{x}$. Assume, by the way of
  contradiction, that \Unjp holds, so $k_{x}$ is below the \rhs of the
  inclusion.  Since the only \mjc $D$ of $k_{x}$ such that
  $D \refines \set{k_{x}}$ is $\set{k_{x}}$, either
  $k_{x} \leq \rd[y] \vee \ld[z] \vee w$ or
  $k_{x} \leq \ld[y] \vee \rd[z] \vee w$; let us assume that the first
  case holds.  We have then
  $k_{x} \leq \rd[y] \vee k_{z} \vee \bv (C_{x} \setminus
  \set{k_{y},k_{z}}) = \rd[y] \vee \bv (C_{x} \setminus
  \set{k_{y}})$.
  Considering that $\rd[y] \leq k_{y}$, Lemma~\ref{lemma:elofmjc}
  implies that
  $k_{y} = \rd[y] = (y_{0} \land y_{1}) \vee (y_{0} \land
  y_{2})$.
  Since $k_{y}$ is \jirr $k_{y} = y_{0} \land y_{i}$ for some
  $i \in \set{1,2}$. Yet this is not possible, as such relation
  implies that $C_{y,i}$ is a \jc of $k_{y}$; considering that
  $C_{y,i}$ is a proper subset of the \mjc $C_{y}$, this contradicts
  the minimality of $C_{y}$.  If
  $k_{x} \leq \ld[y] \vee \rd[z] \vee w$, then we get to a similar
  contradiction. Whence, $k_{x}$ is not below the \rhs of \Unjp, which
  therefore fails.  \qed
\end{proof}

It is worth noticing that the statement ``\emph{every \mjc contains
  exactly one \njp element}'' is not definable by equations: for
$A = D = \set{0,1}$, there is a sublattice of $\R(D,A)$ which fails
this property.

While Theorem~\ref{thm:corrunjp} gives a semantic characterization of
\Unjp, we might also wish to measure its power at the syntactic level.
Theorem~\ref{thm:unjp} and Corollary~\ref{cor:unjp} illustrate the
deductive strength of \Unjp, by pinpointing an infinite set of its
consequences.
\begin{theorem}
  \label{thm:unjp}
  If $s_{\ell} = s_{\rho}$ and $t_{\ell} =
  t_{\rho}$ are equations valid on distributive lattices, then 
  the \identity
  \begin{align*}
    \smalllhs[40mm]{
      (x \land (s_{\ell} \vee t_{\ell} \vee w)) 
      \vee (x \land (s_{\rho} \vee t_{\rho} \vee w)) } & \\
    & = (x \land (s_{\rho} \vee t_{\ell} \vee w)) \vee (x \land (s_{\ell} \vee t_{\rho} \vee w))
  \end{align*}
  is derivable from \Unjp and general lattice axioms.
\end{theorem}
\begin{proof}
  For a lattice term $s$, let $\DNF(s)$ be its disjunctive normal
  form.  Recall that we can obtain $\DNF(s)$ from $s$ by means of a
  sequence $s = s_{0},\ldots ,s_{n} = \DNF(s)$ where, for each $i =
  0,\ldots, n-1$, $s_{i +1}$ is obtained from $s_{i}$ by one
  application of the distributive law at the toplevel of the term, and
  by general lattice axioms.  Thus, for two lattice terms
  $s^{1},s^{2}$, let $s^{1}_{i}$, $i = 0,\ldots ,n$, and $s^{2}_{j}$,
  $j = 0,\ldots ,m$ be the sequences leading to the respective normal
  forms.

  For $i =0,\ldots ,n$ and $j = 0,\ldots ,m$, let now $t_{i,j} = x
  \land (s^{1}_{i} \vee s^{2}_{j} \vee w)$. Using \Unjp and general
  lattice axioms, we can compute as follows:
  \begin{align*}
    t_{0,0}  = t_{1,0} \vee t_{0,1} & = t_{2,0} \vee t_{1,1} \vee
    t_{0,2} = \ldots \\
    & = \bv_{j=0,\ldots, m} t_{n,j} \vee \bv_{i=0,\ldots ,n} t_{i,m} \overset{?}{=}
    t_{n,0} \vee t_{0,m}\,,
  \end{align*}
  where only the last equality needs to be justified.  Notice that the
  relation $s^{k}_{i+1} \leq s^{k}_{i}$ holds in every lattice. Whence
  we have $s^{k}_{i'}\leq s^{k}_{i}$ when $i < i'$, and both $t_{n,j}
  \leq t_{n,0}$ and $t_{i,m} \leq t_{i,0}$. It follows that the
  indexed join at the last line evaluates to $t_{n,0} \vee t_{0,m}$.
  We have derived, up to now, the identity
  \begin{align*}
    x \land (s^{1} \vee s^{2} \vee w) & = (x \land (\DNF(s^{1}) \vee
    s^{2} \vee w)) \vee (x \land (s^{1} \vee \DNF(s^{2}) \vee w))
  \end{align*}
  for every pair of lattice terms $s^{1}$ and $s^{2}$.

  Let us call co-clause a conjunction of variables. By using lattice
  axioms only, we can suppose that, within $\DNF(t)$, there are no
  repeated literals in co-clauses and that no co-clause subsumes
  another. Under this assumption, we have that an identity $s_{\ell} =
  s_{\rho}$ holds in all distributive lattices if and only if
  $\DNF(s_{\ell})$ is equal to $\DNF(s_{\rho})$.
  Whence, to derive the statement of the Theorem, we can compute as
  follows:
  {\allowdisplaybreaks
  \begin{align*}
    \smalllhs{(x \land (s_{\ell} \vee t_{\rho} \vee w)) \vee (x \land
      (s_{\rho}
      \vee t_{\ell}  \vee w))} \\
    & = (x \land (\DNF(s_{\ell}) \vee t_{\rho} \vee w)) \vee (x \land
    (s_{\ell} \vee
    \DNF(t_{\rho}) \vee w)) \\
    & \qquad \vee (x \land (\DNF(s_{\rho}) \vee t_{\ell} \vee w)) \vee
    (x \land
    (s_{\rho} \vee \DNF(t_{\ell}) \vee w)) \\
    & = (x \land (\DNF(s_{\rho}) \vee t_{\rho}) \vee w) \vee (x \land
    (s_{\ell} \vee
    \DNF(t_{\ell}) \vee w)) \\
    & \qquad \vee (x \land (\DNF(s_{\ell}) \vee t_{\ell} \vee w)) \vee
    (x \land
    (s_{\rho} \vee \DNF(t_{\rho})  \vee w)) \\
    & = (x \land (\DNF(s_{\ell}) \vee t_{\ell} \vee w)) \vee (x \land
    (s_{\ell} \vee
    \DNF(t_{\ell}) \vee w)) \\
    & \qquad \vee (x \land (\DNF(s_{\rho}) \vee t_{\rho} \vee w)) \vee
    (x \land (s_{\rho} \vee \DNF(t_{\rho}) \vee w)) \tag*{---where we
      have permuted the order of the four joinands}
    \\
    & = (x \land (s_{\ell} \vee t_{\ell} \vee w)) \vee (x \land
    (s_{\rho} \vee t_{\rho} \vee w))\,.  \tag*{\qed}
  \end{align*}
}
\end{proof}

In \cite{LitakMHjlamp} two equations were shown to hold on
relational lattices. One of them is \AxRLtwo that we describe
next. Set
\begin{align*}
    \lcd[u] := (u_{0} \vee u_{1}) \land (u_{0} \vee u_{2})\,, \quad
    \rcd[u] := u_{0} \vee (u_{1} \land u_{2})\,,
\end{align*}
the equation is
\begin{align}
  \label{eq:AxRL2}
  \tag{RL2}
  x \land (\lcdy \vee \lcdz) 
  & \leq (x \land (\rcd \vee \lcdz)) \vee (x \land (\lcdy \vee \rcdz))
\end{align}

\begin{corollary}
  \label{cor:unjp}
  If $s_{\ell} = s_{\rho}$ and $t_{\ell} = t_{\rho}$ are equations
  valid on distributive lattices, then the inequation
  \begin{align}
    \label{eq:genAXRL2}
    (x \land (s_{\ell} \vee t_{\ell} \vee w)) 
    & \leq (x \land (s_{\rho} \vee t_{\ell} \vee w)) \vee (x \land (s_{\ell} \vee t_{\rho} \vee w))
  \end{align}
  is derivable from \Unjp. In particular \AxRLtwo is derivable from
  \Unjp.
\end{corollary}
The Corollary follows from the Theorem and from the fact that $x \leq
x \vee y$. In order to derive \AxRLtwo from \eqref{eq:genAXRL2} (if we
do not include the bottom constant $\bot$ as part of the signature of
lattice theory), we instantiate $s_{\ell} := \lcd[y]$, $s_{\rho} :=
\rcd[y]$, $t_{\ell} := \lcd[z]$, $t_{\rho} := \rcd[z]$, and $w : =
\rcd[z]$.

\medskip

It can be shown that \Unjp is not derivable from \AxRLtwo---mainly due
to the role of the variable $w$ in the \Unjp. 
The construction of a lattice $L$ satisfying
\AxRLtwo but failing \Unjp proceeds via the construction of its
OD-graph $\langle J(L), \leq ,\mcovered\rangle$.  Due to the
consistent number of variables in the two equations, an automated tool
such as \Macef \cite{prover9-mace4} could not help finding a
countermodel. Similarly, automated tools such as \Provern and
\Waldmeister \cite{prover9-mace4,waldmeister} were of no help to show
that \AxRLtwo is a consequence of \Unjp.

\medskip 

Natural questions---e.g. decidability---may be raised concerning the
equational theory of \Unjp.
Since we can give an easy semantic proof that an equation of the form
\eqref{eq:genAXRL2} holds on finite lattices (or \pperfect) satisfying
\Unjp, a reasonable conjecture is that this theory has some sort of
finite model property. Yet, proving this might not be immediate, since
the variety of lattices satisfying \Unjp is not locally finite (i.e.,
not every finitely generated lattice satisfying \Unjp is finite). The
construction used in \cite[Proposition 7.5]{San09:duality} may be used
to argue that the lattice freely generated in this variety by three
generators is infinite.

\section{Symmetry and \PC}
\label{sec:othereqs}

Due to its syntactic shape \Unjp falls in a class of inclusions
described in \cite[Section 8]{San09:duality} that admit a
correspondent property in the OD-graph. Here, the meaning of the word
correspondent is analogous to its use in modal logic, where some
formulas might be uniformly valid in a frame if and only if the frame
satisfies a correspondent first order property. Thus
Theorem~\ref{thm:corrunjp} is not completely unexpected. A more
surprising result comes from considering the three equations below,
that fall outside the syntactic fragment described in
\cite{San09:duality}; a strengthening of Lemma~\ref{lemma:elofmjc}
(Lemma~\ref{lemma:subsofmjc} to follow) allows to characterize the
OD-graphs of \pperfect lattices satisfying \Unjp and these equations,
see Theorem~\ref{thm:characterization}.
\begin{align}
  \label{eq:N1}
  \tag{SymPC}
  x \land (y \vee z) & \leq \\
  & \smallrhs[70mm]{(x \land (y \vee (z \land (x \vee y)))) \vee (x
    \land (z \vee (y \land (x \vee z))))}
  \notag
  \\
  \tag{VarRL1}
  \label{eq:MyAxRL1}
  x \land ((y \land z) \vee (y \land x) \vee  (z \land x))
  & \leq (x \land y) \vee (x \land z) \\
  \tag{RMod}
  \label{eq:RMod}
  x \land ((x \land y) \vee \ld[z])
  & \leq (x \land ((x \land y) \vee \rd[z])) \vee (x \land \ld[z])
\end{align}
Let us first illustrate the way in which these equations hold in
relational lattices. In particular, the proof shall illustrate the
crucial role played by symmetry and \PC---i.e., condition
\eqref{cond:BC}---of the ultrametric space $(\AD,\d)$.
\begin{theorem}
  \label{thm:eqnshold}
  The inclusions \SymBC, \MyAxRLone, \RMod hold in relational lattices.
\end{theorem}
\begin{proof}
  \SymBC.  Let $k$ be a \jirr element below $x \land (y \vee z)$. If
  $k$ is \jp, then $k$ is also below $(x \land y) \vee (x \land z)$,
  whence it is below the \rhs of this inclusion.  Therefore, let $k$
  be \njp, so $k = \JP{f}$ for some $f \in \AD$; by
  \eqref{eq:jirrmjc}, let $g \in \AD$ be such that
  $\JP{\delta(f,g)} \cup \set{\JP{g}} \refines \set{y,z}$.  Let us
  suppose first that $\JP{g} \leq z$. Since
  $\JP{\delta(f,g)} \refines \set{y ,z}$, using \PC we can find $h$
  such that $\JP{\delta(f,h)} \refines \set{y}$ and
  $\JP{\delta(h,g)} \refines \set{z}$.  It follows that
  $\JP{h} \leq \bv \JP{\delta(h,g)} \vee \JP{g} \leq z$; moreover,
  since $\JP{f} \leq x$, $\delta(h,f) = \delta(f,h)$, and
  $\JP{\delta(f,h)} \refines \set{y}$, then
  $\JP{h} \leq \JP{\d(h,f)} \vee \JP{f} \leq x \vee y$. Consequently,
  we have $\JP{h} \leq z \land (x \vee y)$ and, considering that
  $\JP{\delta(f,h)} \refines \set{y}$, we have
  $\JP{f} \leq x \land (y \vee (z \land (x \vee y)))$.
  If $\JP{g} \leq y$, then we similarly deduce that
  $\JP{f} \leq x \land (z \vee (y \land (x \vee z)))$. In both cases,
  $\JP{f}$ is below the \rhs of this inclusion.

  \medskip

  \MyAxRLone. 
  Let $k$ be below the \lhs of this inclusion. If $k$ is \jp, then it
  is below the \rhs of this inclusion as well. Otherwise $k = \JP{f}$
  is \njp and
  $\JP{\d(f,g)} \cup \set{\JP{g}} \refines \set{y \land z, y \land x,z
    \land x}$ for some $g \in \AD$.

  Since $\JP{g} \leq r$ for some
  $r \in \set{y \land z, y \land x,z \land x}$, we consider three
  cases; by \PC we can also assume that $\JP{g}$ is the only element
  of $\JP{\d(f,g)} \cup \set{\JP{g}}$ below $r$---since if
  $\d(f,g') \refines \set{y \land z, y \land x,z \land x} \setminus
  \set{r} $
  and $\d(g',g) \refines \set{r}$, then $g' \leq r$. Also, the last
  two cases, $g \leq y \land x$ and $g \leq z \land x$, are symmetric
  in $y$ and $z$, so that we consider among them the second-to-last
  only.

  Suppose firstly that $\JP{g} \leq y \land z$.  Then, from
  $\JP{\d(g,f)} = \JP{\d(f,g)} \refines \set{x\land y , x \land z }
  \refines \set{x}$ and $\JP{f} \leq x$, we deduce $\JP{g} \leq x$;
  whence $\JP{g} \leq x \land y$ and $\JP{f} \leq (x \land z) \vee
  (x\land y)$.
  
  Suppose next that $\JP{g} \leq x \land y$.  By \PC, let $h$ be such
  that $\JP{\d(f,h)} \refines \set{x \land z}$ and
  $\JP{\d(h,g)} \refines \set{y \land z}$. We deduce then
  $\JP{h} \leq y$ from $\JP{\d(h,g)} \refines \set{y}$ and
  $\JP{g} \leq y$, and $\JP{h} \leq x$, from
  $\JP{\d(h,f)} = \JP{\d(f,h)} \refines \set{x}$ and $\JP{f} \leq
  x$.
  Thus $\JP{h} \leq x \land y$ and
  $\JP{f} \leq (x \land z) \vee (x\land y)$.

  \medskip

  \RMod. 
  Let $k$ be a \jirr below the \lhs of this inclusion. If $k$ is \jp,
  then $k$ is below $x \land ( (x \land y) \vee \rd[z])$. Otherwise
  $k = \JP{f}$ and, for some $g \in \AD$,
  $\JP{\d(f,g)} \cup \set{\JP{g}} \refines \set{x \land y,\ld[z]}$. If
  $\JP{g} \leq x \land y$, then all the elements that are not below
  $x \land y$ are below $\ld[z]$ and \jp, whence they are below
  $\rd[z]$. It follows that
  $\JP{f} \leq x \land ( (x \land y) \vee \rd[z])$.  Otherwise
  $\JP{g} \leq \ld[z]$ and, by \PC, we can also assume that $g$ is the
  only element below $\ld[z]$, so
  $\JP{\d(f,g)} \refines \set{x \land y}$. It follows then that
  $\JP{\d(g,f)} \cup \set{\JP{f}} = \JP{\d(f,g)} \cup \set{\JP{f}}
  \refines \set{x}$,
  $g \leq x$, whence $\JP{g} \leq x \land \ld[z]$. Consequently,
  $\JP{f} \leq (x \land y) \vee (x \land \ld[z]) \leq (x \land ((x
  \land y)\vee \rd[z])) \vee (x \land \ld[z])$.  \qed
\end{proof}

In \cite{LitakMHjlamp} a second inclusion was shown to hold on
relational lattices:
\begin{align}
  \label{eq:AxRL1}
  \tag{RL1}
  x \land ((y \land (z \vee x)) \vee (z \land (y \vee x))) &
  \leq (x \land y) \vee (x \land z)
\end{align}
The same kind of tools used in the proof of Theorem~\ref{thm:eqnshold} can be used
to argue that this inclusion holds on relational lattices. 
The reader will have noticed the similarity of \MyAxRLone with
\AxRLone. As a matter of fact, \MyAxRLone was suggested when trying to
derive \AxRLone from \Unjp and the other equations as in the following
Proposition.
\begin{proposition}
  \AxRLone is a consequence of \Unjp, \RMod and \MyAxRLone.
\end{proposition}
\begin{proof} 
  Using \eqref{eq:Unjp} and considering that $y \land z \leq z \land
  (y \vee x)$, we have:
  \begin{align*}
    \smalllhs[15mm]{x \land ((y \land (z \vee x)) \vee (z \land (y
      \vee
      x)))} \\
     & = (x \land ((y \land x) \vee (z \land (y \vee x)))) \vee (x
    \land ((y \land (z \vee x)) \vee (z \land x)))) \,.
  \end{align*}
  Using now \RMod and considering that $x \land z \leq y \vee x$, we
  compute as follows:
  \begin{align*}
    \smalllhs[30mm]{x \land ((y \land x) \vee (z \land (y \vee x)))} &
    \\
    =\quad & (x \land ((y \land x) \vee (z \land y) \vee (z \land x)))
    \vee (x \land z \land (y \vee x)) \\
    = \quad & (x \land ((y \land x) \vee (z \land y) \vee (z \land
    x))) \vee (x \land z) \\
    = \quad & x \land ((y \land z) \vee (y \land x) \vee (z \land x))\,.
  \end{align*}
  Considering the symmetric role of $y$ and $z$, we obtain:
  \begin{align*}
    x \land ((y \land (z \vee x)) \vee (z \land (y \vee x))) 
    = \quad & x \land ((y \land z) \vee (y \land x) \vee (z \land
    x))\\
    = \quad & (x \land y) \vee (x \land z)\,,
    \tag*{by \eqref{eq:MyAxRL1}. \qquad\qed}
  \end{align*}
\end{proof}

We present now what we consider our strongest result in the study of
the equational theory of relational lattices. To this end, let us
denote by \Axioms the (set composed of the) four equations \Unjp,
\MyAxRLone, \RMod and \SymBC.  Also, given that we restrict to
lattices satisfying \Unjp, and considering the characterization given
with Theorem~\ref{thm:corrunjp}, it is convenient to introduce the
notation $\DSTEP{k_{0}}{C}{k_{1}}$ for the statement
\emph{$k_{0},k_{1} \in J(L)$, $k_{1}$ is \njp, $k_{1} \not\in C$, and
  $k_{0} \mcovered C \cup \set{k_{1}}$}.
\begin{theorem}
  \label{thm:atomisticcharacterization}
  Let $L$ be a finite atomistic lattice.  Then 
  $L \models \Axioms$ if and only if every nontrivial \mjc
  contains exactly one \njp element and, moreover, the following
  properties hold in the OD-graph:
  \begin{align}
    \label{prop:SymAtomistic}
    \myitem{ii} \text{If $\DSTEP{k_{0}}{C}{k_{1}}$, then
      $\DSTEP{k_{1}}{C}{k_{0}}$.}
    \\
    \notag \myitem{iii} \text{If
      $\DSTEP{k_{0}}{C_{0} \sqcup C_{1}}{k_{2}}$, then
      $\DSTEP{k_{0}}{C_{0}}{k_{1}}$ and
      $\DSTEP{k_{1}}{C_{1}}{k_{2}}$, 
        for some $k_{1} \in \Ji(L)$.} 
\end{align}
\end{theorem}
Given Theorem~\ref{thm:atomisticcharacterization}, it becomes tempting
to look for a representation Theorem. Given a \pperfect atomistic
lattice satisfying the above four equations, we would like to define
an ultrametric space on the set of \njp elements with distance valued
on the powerset of the \jp ones, and then argue that the lattice
constructed via the standard action, defined in \eqref{def:distance},
is isomorphic to the given lattice. Unfortunately, this idea does not
work, since if we try to set $\d(k_{0},k_{1}) = C$ whenever
$\DSTEP{k_{0}}{C}{k_{1}}$, this might be ill defined since the
implication ``$\DSTEP{k_{0}}{C}{k_{1}}$ and $\DSTEP{k_{0}}{D}{k_{1}}$
implies $C = D$'' might fail. Moreover, there is no equation nor
quasiequation enforcing this, as an immediate consequence of the next
Proposition.
\begin{proposition}
  There is an atomistic sublattice of $\R(\set{0,1},\set{0,1})$ which
  does not arise from an ultrametric space.
\end{proposition}

Theorem~\ref{thm:atomisticcharacterization} is a consequence of a more
general Theorem, to be stated next, characterizing the OD-graphs of
\pperfect lattices in the variety axiomatized by \Axioms.  While the
conditions stated next may appear quite complex, they are the ones to
retain if we aim at studying further the theories of relational
lattices by duality---e.g., a sublattice of a relational lattice need
not be atomistic.
\begin{theorem}
  \label{thm:characterization}
  A \pperfect lattice 
  belongs to the variety axiomatized by \Axioms if and only if every
  minimal join-cover contains at most one \njp element and, moreover,
  the following properties hold in its
  OD-graph: 
  \begin{align}
    \tag{{\prop}\ref{eq:MyAxRL1}}
    \label{prop:MyAxRL1}
    \myitem{i} \text{If $k_{0} \mcovered C$, then there exists at most
      one $c \in C$ with $c \leq k_{0}$.} \\
    \myitem{iv} \tag{{\prop}RMod}
    \label{prop:RMod}
    \text{If $\DSTEP{k_{0}}{C}{k_{1}}$, then no element of $C$ is
      below $k_{0}$.}  \\ 
    \notag
    \myitem{i} \text{if $k \mcovered C_{0} \sqcup C_{1}$ with
      $C_{0},C_{1}$ non-empty, then for some 
      $k' \in \Ji(L)$,} \\
    \notag
    &\quad \text{either $k\mcovered \set{k'} \sqcup C_{1}$,
      $k' \mcovered C_{0}$, and $k' \leq \bv C_{1} \vee
      k$,} \\
    &\qquad\quad\text{or $k\mcovered C_{0} \sqcup \set{k'}$,
      $k' \mcovered C_{1}$, and $k' \leq \bv C_{0} \vee k$.}
    \tag{{\prop}StrongSymPC}
    \label{prop:PSSymBC} 
\end{align}
\end{theorem}

Let us notice that the conditions
\begin{align}
    \notag
    \myitem{iii} \text{If $\DSTEP{k_{0}}{C_{0} \sqcup C_{1}}{k_{2}}$ then, for some $k_{1} \in \Ji(L)$,}\\
    & \qquad\qquad \text{$\DSTEP{k_{0}}{C_{0}}{k_{1}}$,
      $\DSTEP{k_{1}}{C_{1}}{k_{2}}$, and $k_{1} \leq \bv C_{0} \vee
      k_{0}$.} 
    \tag{{\prop}SymPC}
    \label{prop:SymBC} \\
  \tag{{\prop}Sym}
  \label{prop:Sym}
  \myitem{ii} \text{\em If $\DSTEP{k_{0}}{C}{k_{1}}$, then
    $k_{1}\leq \bv C \vee k_{0}$}  \hspace{45mm}
\end{align}
follow from the above properties.  On atomistic \pperfect lattices the
last condition is equivalent to \eqref{prop:SymAtomistic}.

\begin{lemma}
  \label{lemma:propSymAtomistic}
  If $L$ is a \pperfect lattice with $L \models \Unjp$ and whose
  OD-graph satisfies \PSSymBC and \PRMod, then
  \PSymBC holds as well. 
\end{lemma}
\begin{proof}
  Let $k_{1} \mcovered C_{0} \sqcup C_{1} \sqcup \set{k_{2}}$ with
  $k_{2} \in \Ji(L)$ and \njp, and use \PSSymBC to find $k_{1}$
  such that either (i)
  $k_{0} \mcovered \set{k_{1}} \sqcup C_{1} \sqcup \set{k_{2}}$ and
  $k_{1} \leq k_{0} \vee \bv C_{1} \vee k_{2}$, or (ii)
  $k_{0} \mcovered C_{0} \sqcup \set{k_{1}}$ and
  $k_{1} \leq k_{0} \vee \bv C_{0}$.
  Let us argue, by contradiction, that (i) cannot arise. By \Unjp,
  $k_{1}$ is \jp, whence the relation $k_{1} \leq k_{0} \vee \bv
  C_{1}$ yields $k_{1} \leq k_{0}$. This, however, contradicts \PRMod.
  \qed
\end{proof}

We close this section by proving Theorem~\ref{thm:characterization}.
To this end, we need a generalization of Lemma~\ref{lemma:elofmjc}.
As the refinement relation is an extension to subsets of the order
relation, the relation $\wrefines$, defined next, can be considered as
an extension to subsets of the \mjc{ing} relation.
\begin{definition}
  Let $L$ be a \pperfect lattice and let $X, Y \subseteq \Ji(L)$ be
  antichains. Put $X \wrefines Y$ if $X \refines \set{\bv Y}$ and
  $y \in C_{x_{y}}$ for some $x_{y} \in X$, for each $y \in Y$ and
  whenever $\set{C_{x} \mid x \in X}$ is a family of coverings of the
  form $x \mcovered C_{x} \refines Y$.
\end{definition}
\begin{lemma}
  \label{lemma:subsofmjc}
  Let $L$ be a \pperfect lattice and let
  $j \mcovered C_{0} \sqcup C_{1}$.  Suppose that
  $\bv X \leq \bigvee C_{0}$ and $j \leq \bv X \vee \bv C_{1}$. Then
  there exists a \mjc of the form $j \mcovered D_{0} \sqcup C_{1}$
  with $D_{0} \refines X$ and $D_{0} \wrefines C_{0}$.
\end{lemma}

While the proof that a \pperfect lattice whose OD-graphs satisfies
those properties essentially mimics the proof of
Theorem~\ref{thm:eqnshold},
we prove instead the converse direction through a series of
Lemmas.
\begin{lemma}
  \label{lemma:MyAxRL1}
  If \eqref{eq:MyAxRL1} holds on a \pperfect lattice, then its
  OD-graph satisfies \eqref{prop:MyAxRL1}.
\end{lemma}
\begin{proof}
  Suppose $C = \set{k_{1}} \sqcup \set{k_{2}} \sqcup D$ with
  $k_{0} \mcovered C$ and $k_{1},k_{2} \leq k_{0}$. Let $x := k_{0}$,
  $y := k_{1} \vee \bv D$, $z := k_{2} \vee \bv D$. Then
  $k_{1} \leq x \land y$, $k_{2} \leq x \land z$ and
  $\bv D \leq y \land z$, whence the \lhs of \MyAxRLone evaluates to
  $k_{0}$, which is therefore below the \rhs of this inclusion. It
  follows that either $k_{0} \leq y$, or $k_{0} \leq z$, in both cases
  contradicting the fact that $C$ is a \mjc.  \qed
\end{proof}

The inclusion 
\begin{align}
  \tag{Sym}
  \label{eq:Sym}
  x \land (y \vee \ld[z]) & \leq (x \land (y \vee \rd[z])) \vee (x
  \land (y \vee (\ld[z]\land (y \vee x)))\,,
\end{align}
is derivable from \Unjp, \None and \RMod. It can also be shown that
\RMod is a consequence of \Sym.

\begin{lemma}
  \label{lemma:sym1}
  If \Sym holds in a \pperfect lattice, then its OD-graph satisfies
  \PSym.
\end{lemma}
\begin{proof}
  Suppose that the inclusion holds and let
  $\DSTEP{k_{0}}{C}{k_{1}}$. Since $k_{1}$ is \njp, there exists a
  \mjc $k_{1} \mcovered D$ which we can partition into two non empty
  subsets $D_{1}$ and $D_{2}$.
  Let now $x := k_{0}$, $y := \bv C$, $z_{0} := k_{1}$,
  $z_{1} = \bv D_{1}$, $z_{2} = \bv D_{2}$. Then, the \lhs of the
  inclusion evaluates to $k_{0}$, which therefore is below the
  \rhs. Considering that $x$ is $k_{0}$ and that the unique \mjc of
  $k_{0}$ whose elements are all below $k_{0}$ is $\set{k_{0}}$,
  it follows that either $k_{0} \leq y \vee \rd[z]$ or
  $k_{0} \leq y \vee (\ld[z] \land (y \vee x))$.
  
  Argue that $\rd[z] < \ld[z]$, since $k_{1}$ is \jirr, whence by
  Lemma~\ref{lemma:elofmjc}, $\set{y,\rd[z]}$ is not a cover of
  $k_{0}$, excluding the first case.  Therefore
  $\set{y,\ld[z] \land (y \vee x)}$ is a cover of $k_{0}$, whence, by
  Lemma~\ref{lemma:elofmjc}, $k_{1} = \ld[z] \land (y \vee x)$,
  showing that $k_{1} \leq y \vee x = \bv C \vee k_{0}$ and proving
  the statement.  \qed
\end{proof}

\begin{lemma}
  \label{lemma:Cnbelow}
  \label{lemma:PRMOD}
  If $L$ is a \pperfect lattice such that $L \models \Axioms$, then
  \PRMod holds in its OD-graph.
\end{lemma}
\begin{proof}
  Let $\DSTEP{k_{0}}{C}{k_{1}}$ and put $C = C_{0} \sqcup C_{1}$ with
  $C_{0} \refines \set{k_{0}}$ and $c \not\leq k_{0}$ for each element
  $c \in C_{1}$.  As \Sym whence \PSym hold, we have
  $k_{1} \leq \bv C_{1} \vee \bv C_{0} \vee k_{0}$.  

  We consider next equation \RMod. Put $x := k_{0}$, $y := \bv C_{0}$,
  $z_{0} := \bv C_{1} \vee k_{1}$,
  $z_{1} := \bv C_{1} \vee \bv C_{0}$, $z_{2} := k_{0}$.  From
  $k_{1} \leq \bv C_{1} \vee \bv C_{0} \vee k_{0}$, we get
  $z_{0} \land (z_{1} \vee z_{2}) = z_{0}$. Whence, the \lhs of \RMod
  evaluates to $k_{0}$ so $k_{0}$ is below the \rhs of \RMod.
  Considering that $\set{k_{0}}$ is the unique \mjc of $k_{0}$ whose
  elements are all below $k_{0}$, it follows that either
  $k_{0} \leq z_{0} \land (z_{1} \vee z_{2})$ or
  $k_{0} \leq y \vee (z_{0} \land z_{1}) \vee (z_{0} \land z_{2}) $.

  As $k_{0} \not\leq \bv C_{1} \vee k_{1} = z_{0}$, it follows that $k_{0}
  \leq y \vee (z_{0} \land z_{1}) \vee (z_{0} \land z_{2}) $. We can
  use then Lemma~\ref{lemma:subsofmjc} to deduce that
  $k_{0}$ has a \mjc of the form $k_{0} \mcovered C_{0} \sqcup D$,
  with $D \refines \set{z_{0} \land z_{1},z_{0} \land z_{2}}
  \refines \set{z_{1},z_{2}}$. If all the elements of $D$ are
  below $z_{1} = \bv C_{0} \vee \bv C_{1}$, then
  \begin{align*}
    k_{0} & \leq \bv C_{0} \vee \bv D \leq \bv C_{0} \vee \bv C_{1}\,,
  \end{align*}
  contradicting the minimality of
  $k_{0} \mcovered C_{0} \sqcup C_{1} \sqcup \set{k_{1}}$.  Therefore,
  at least one element of $D$ is below $z_{2} = k_{0}$. If
  $C_{0} \neq \emptyset$, then in the \mjc $C_{0} \sqcup D$ there are
  at least two elements that are below $k_{0}$. This however
  contradicts \PMyAxRLone, whence \MyAxRLone.  We have, therefore,
  $C_{0} = \emptyset$.  \qed
\end{proof}

\begin{lemma}
  \label{lemma:almost}
  If $L$ is a \pperfect lattice with $L \models \Axioms$, then
  \PSSymBC holds in is OD-graph.
\end{lemma}
\begin{proof}
  By Lemmas~\ref{lemma:MyAxRL1} and \ref{lemma:PRMOD}, \PRMod and
  \PMyAxRLone hold in the OD-graph.

  Let $x := k_{0}$, $y := \bv C_{0}$, $z := \bv C_{1}$. Then the \lhs
  of \SymBC evaluates to $k_{0}$ which is therefore below the
  \rhs. Thus, by Lemma~\ref{lemma:subsofmjc}, there is a \mjc of the form
  $k_{0} \mcovered D_{0} \sqcup D_{1}$ with either (i)
  $D_{0} \wrefines C_{0}$, $D_{0} \refines \set{y \land (x \vee z)}$,
  and $D_{1} = C_{1}$, or (ii) $D_{0} = C_{0}$,
  $D_{1} \wrefines C_{1}$, and $D_{1} \refines \set{z \land (x \vee y)}$.

  W.l.o.g. we can suppose that (i) holds.  From
  $D_{0} \refines \set{y \land (x \vee z)} \refines \set{x \vee z} =
  \set{k_{0} \vee \bv C_{1}}$,
  we can argue as follows. We notice first that if an element of
  $D_{0}$ is \jp, then it is either below $k_{0}$ or below some
  $c \in C_{1}$; since $D_{0} \sqcup C_{1}$ is an antichain, this
  element is below $k_{0}$.  Therefore, if all the elements of $D_{0}$
  are \jp, then, by \PMyAxRLone, $D_{0} = \set{k'}$.
  Otherwise, there exists a \njp element $k'$ in $D_{0}$ and, by
  \Unjp, this is the only \njp in $D_{0}$.  Write
  $D_{0} = \set{k'} \sqcup E$, then every element of $E$ is \jp and,
  as seen before, we need to have $E \refines \set{k_{0}}$. Then \PRMod
  enforces $E_{0} = \emptyset$ and $D_{0}= \set{k'}$.
  In both cases,  the relation $\set{k'} = D_{0} \wrefines
  C_{0}$ yields $k' \mcovered C_{0}$. 
  \qed
\end{proof}

Finally, in order to understand the structure of finite lattices in
the variety of axiomatized by \Axioms,
let $\Jp(L)$ denote the set of \jp elements of $L$ and consider the
following property:
\begin{align*}
  \myitem{iv} 
  \text{\em If $k_{0} \mcovered C$ and
    $C \subseteq \Jp(L)$, then $c_{0} \leq k_{0}$ for some
    $c_{0} \in C$} 
  \tag{{\prop}JP}
  \label{prop:JP} 
\end{align*}
The next Lemma ensures the existence of a \njp element in a cover in
finite atomistic lattices, as stated in
Theorem~\ref{thm:atomisticcharacterization}.
\begin{lemma}
  \label{prop:pjp}
  If a \emph{finite} lattice $L$ satisfies \Axioms, then \PJP holds in
  its OD-graph. In particular, if $L$ is atomistic, then
  $k_{0} \mcovered C$ implies that $k_{1} \in C$ for some \njp
  $k_{1}$.
\end{lemma}
It can be shown that the finiteness assumption in Lemma~\ref{prop:pjp}
is necessary.

\section{Conclusions and further directions}
\label{sec:conclusions}

\subsubsection{Some undecidable problems.}
Our main result, Theorem~\ref{thm:atomisticcharacterization},
characterizes the OD-graphs of finite atomistic lattices satisfying
\Axioms as structures similar to frames for the commutator logic
$[\Sfive]^{n}$, the multimodal logic with $n$ distinct pairwise
commuting $\Sfive$ modal operators, see \cite{Kurucz2007}. We
exemplify next how to take advantage of such similarity and of the
existing theory on combination of modal logics, to deduce
undecidability results. As this is not the main goal of the paper, we
delay a full exposition of these ideas to an upcoming set of notes.
 
An $[\Sfive]^{n}$ frame is a structure $\FF = (F,R_{1},\ldots ,R_{n})$
where each $R_{i}$ is an equivalence relation on $F$ and, moreover,
the confluence property holds: \emph{if $i \neq j$, $xR_{i}y$ and
  $xR_{j}z$, then $y R_{j}w$ and $zR_{i} w$ for some $w \in F$}.  A
particular class of $[\Sfive]^{n}$ frames are the universal
$\Sfive^{n}$-products, those of the form
$\U = (F,R_{1},\ldots ,R_{n})$ with
$F = X_{1} \times \ldots \times X_{n}$ and
$(x_{1},\ldots ,x_{n})R_{i}(y_{1},\ldots ,y_{n})$ if and only if
$x_{j} = y_{j}$ for each $j \neq i$.

For a frame $\FF = (W, R_{1},\ldots ,R_{n})$ and
$X \subseteq \set{1,\ldots ,n}$, let us say that $Y \subseteq W$ is
$X$-closed if $w_{0} \in Y$, whenever there is a path
$w_{0}R_{i_{0}}w_{1}\ldots w_{k-1}R_{i_{k}} w_{k}$ with
$\set{i_{0},\ldots ,i_{k}} \subseteq X$ and $w_{k} \in Y$. Then
$X$-closed subsets are closed under intersections, so subsets of
$\set{1,\ldots ,n}$ give rise to closure operators $\pos[X]$ and to an
action as defined in Section~\ref{sec:rellattices}. Let
$\L(\FF) = P(\set{1,\ldots ,n}) \sdp P(W)$ and notice that $\L(\FF)$
is atomistic.  A frame $\FF$ is initial if there is $f_{0} \in F$ such
every other $f \in F$ is reachable from $f_{0}$; it is full if, for
each $i = 1,\ldots ,n$, $R_{i}$ is not included in the identity. If
$\FF$ is initial and full, then $\L(\FF)$ is subdirectly irreducible.
A $p$-morphism is defined as usual in modal logic.  The key
observation leading to undecidability is the following statement.
\begin{theorem}
  There is a surjective $p$-morphism from a universal
  $\Sfive^{n}$-product frame $\U$ to a full initial frame $\FF$ if and
  only if $\L(\FF)$ embeds in a relational lattice.
\end{theorem}
\begin{proof}[Sketch]
  The construction $\L$ is extended to a contravariant functor, so if
  $\psi : \U \rto \FF$ is a $p$-morpshim, then we have an embedding
  $\L(\psi)$ of $L(\FF)$ into $\L(\U)$. We can assume that all the
  components $X_{1}, \ldots , X_{n}$ of $\U$ are equal, so $X_{i} = X$
  for each $i = 1,\ldots ,n$; if this is the case, then $\L(\U)$ is
  isomorphic to the relational lattice $\R(\set{1,\ldots
    ,n},X)$.  

  The converse direction is subtler. Let $\chi : \L(\FF) \rto \R(A,D)$
  be a lattice embedding; since $\L(\FF)$ is \si, we can suppose that
  $\chi$ preserves bounds; its left adjoint
  $\mu : \R(A,D) \rto \L(\FF)$ is then surjective. Since both
  $\L(\FF)$ and $\R(D,A)$ are generated (under possibly infinite
  joins) by their atoms, each atom $x \in \L(\FF)$ has a preimage
  $y \in \R(D,A)$ which is an atom.  Consider now
  $S_{0} = \set{ f \in \AD \mid \mu(f) \text{ is a \njp atom} }$ and
  make it into a $P(\set{1,\ldots ,n})$-valued ultrametric space by
  letting
  $\d_{S_{0}}(f,g) = \mu(\d(f,g)) \subseteq \set{1,\ldots ,n}$---we
  use here the fact that $\mu$ sends \jp elements to \jp elements.
  $S_{0}$ is shown to be a pairwise complete ultrametric space over
  $\set{1,\ldots ,n}$. We prove that pairwise complete ultrametric
  spaces over a finite set $B$ are in bijection with universal
  $\Sfive^{n}$-product frames, with $n = \card B$.  Then the
  restriction of $\mu$ to $S_{0}$ is a surjective p-morphism from
  $S_{0}$ to (a frame isomorphic to) $\FF$.  \qed
\end{proof}
In view of the following statement, which relies on \cite{HH2001} and
can be inferred from \cite{HHK2002}:
``\emph{
  for $n \geq 3$, it is undecidable whether, given a finite full
  initial frame $\FF$, there is a surjective $p$-morphism from a
  universal $\Sfive^{n}$-product $\U$ to $\FF$'' },
we deduce the following undecidability results, which partially answer
Problem~4.10 in \cite{LitakMHjlamp}.

\begin{corollary}
  It is undecidable whether a finite subdirectly irreducible atomistic
  lattice embeds into a relational lattice.
  Consequently, the quasiequational theory of relational lattices in
  the pure lattice signature is also undecidable.
\end{corollary}

\subsubsection{Comparison with Litak et al. \cite{LitakMHjlamp}.}
We have presented our first contribution to the study of the
equational theory of relational lattices. In \cite{LitakMHjlamp} two
equations in the larger signature with the header constant are
presented as a base for the equational theory of relational
lattices. As mentioned there, the four equations of \Axioms are
derivable from these two equations. Therefore, we can also think of
the present work as a contribution towards assessing or disproving
completeness of these two axiomatizations. 
Yet, we wish to mention here and emphasize some of our original
motivations. Lattice theoretic equations are quite difficult to grasp,
in particular if considered on the purely syntactic side, as done for
example in \cite{LitakMHjlamp}.  Duality theory attaches a meaning to
equations via the combinatorial properties of the dual spaces. This
process is nowadays customary in modal and intuitionistic logic and
gives rise to a well defined area of research, correspondence
theory. Our aim was to attach meaning to the equations of relational
lattices. The answer we provide is, at the present state of research,
via the relevant combinatorial properties, symmetry and \PC.  From
this perspective, the results presented in Section~\ref{sec:othereqs}
undoubtedly need further understanding. In particular it is worth
trying to modularize them, so as to discover equations exactly
corresponding to symmetry or, respectively, to \PC; alternatively,
argue that these equations do not exist. Finally, the present work
opens new directions and challenges for the duality theory developed
in \cite{San09:duality}---of which, we hope we have illustrated the
fruitfulness---including a better understanding of how to generalize
it to the infinite case, new mechanisms by which to devise
correspondence results, natural conjectures concerning equations
having correspondents in finite lattices.

\bibliographystyle{splncs03}
\bibliography{biblio}

\end{document}